\documentclass[12pt]{iopart}

\usepackage{iopams}
\usepackage{graphicx,epsfig,rotate}%,psfig}

\newcommand{\s}{\mathrm{s}}
\newcommand{\q}{\mathrm{Q}}
\renewcommand{\c}{{\hat{\chi}}}

\begin{document}

\title[Scalar-field quintessence by cosmic shear]{Scalar-field quintessence by cosmic
shear:\\CFHT data analysis and forecasts for DUNE\footnote{To appear in Proceedings of IRCAG
06 (Barcelona).}}

\author{C Schimd$^{1}$, I Tereno$^{2,3}$}

\address{${}^{1}$ DAPNIA, CEA Saclay, 91191 Gif-sur-Yvette cedex, France}
\address{${}^{2}$ Argelander-Institut f\"ur Astronomie, Universit\"at Bonn, 53121 Bonn,
Germany}
\address{${}^{3}$ Institut d'Astrophysique de Paris, 98bis bvd. Arago, 75014 Paris, France}

\ead{${}^{1}$carlo.schimd@cea.fr}
\begin{abstract}
A light scalar field, minimally or not-minimally coupled to the metric field, is a
well-defined candidate for the dark energy, overcoming the coincidence problem intrinsic to
the cosmological constant and avoiding the difficulties of parameterizations. We present a
general description of the weak gravitational lensing valid for every metric theory of
gravity, including vector and tensor perturbations for a non-flat spatial metric. Based on
this description, we investigate two minimally-coupled scalar field quintessence models
using VIRMOS-Descart and CFHTLS cosmic shear data, and forecast the constraints for the
proposed space-borne wide-field imager DUNE.
\end{abstract}

%\submitto{...}
%\maketitle

\section{Introduction}

There is an overwhelming evidence that the description of the observed universe cannot rely
only on the assumption of the Copernican principle, on the general relativity, and on the
Standard Model of particle physics, eventually extended to include a dark matter
candidate~\cite{peebles03}. We shall define \emph{dark energy} as anything which represents
the failure of some of these assumptions: \emph{i)} It could be a signature of the dynamical
effect of inhomogeneities, not properly averaged when accounting for the background
dynamics~\cite{inhom}. \emph{ii)} Gravitational interactions (on cosmological scales) could
be not described by the Hilbert-Einstein action, requiring e.g. the inclusion of
higher-order terms as in scalar-tensor theories of
gravity~\cite{DamourEspositoFareseETwands94} or a formulation in more than four dimensions,
as in braneworld scenarios or superstring theories~\cite{morethan4}. \emph{iii)} Dark energy
is an effective ``matter'' field, often dubbed quintessence~\cite{Q0}, not clustering on the
observed scales and possibly coupled to the matter fields~\cite{coupledQ}, provided the weak
equivalence principle is preserved. A combination of these options is the last possibility,
like in extended quintessence scenarios~\cite{EQ0}. One can therefore define \emph{classes}
of models, characterized by definite observational and experimental
signatures~\cite{Uzan06}.

The weak gravitational lensing by large scale structures, or {\it cosmic shear}, is a
geometrical observable which depends on both background evolution and structures
formation~\cite{wlreview}. Therefore it is a promising tool to investigate dark energy
% (in broad sense)
, allowing one to explore in unbiased way the low-redshift universe where dark energy mostly
acts~\cite{HuJain04,wlOQ}. Cosmic shear has been exploited to investigate ordinary
quintessence scenarios, considering a parametrization of the quintessence equation of
state~\cite{HuJain04,wlOQ,wldetect2}.  We move towards ``physics''-inspired models, for the
first time using real data along this strategy~\cite{stum06}. This requires a general
formulation of weak-lensing, which we achieve just supposing that gravitation is described
by a metric theory~\cite{sur04}.

% ==========================================================================================

\section{Geometry of null congruences: cosmic shear}

Every galaxy defines a light bundle, or congruence of null geodesics $ x^\mu(\lambda) =
\bar{x}^\mu(\lambda)+\xi^\mu(\lambda)$ deviating from a fiducial, arbitrary geodesic
$\bar{x}^\mu$ by a displacement $\xi^\mu$, converging at the observer position $O$ (where
$\xi^\mu=0$), and whose tangent vectors $k^\mu\equiv dx^\mu/d\lambda$ are solution of $k_\mu
k^\mu = 0$ and $k^\nu \nabla_\nu k^\mu = 0$ (we take the affine parameter $\lambda$
vanishing in $O$ and increasing toward the past). The shape of the light bundle is described
by a deformation tensor $\mathcal{D}^a_{\;b}$, whose evolution along the fiducial geodesic
(defined by $\bar{k}^\mu$) is deduced from the geodesic deviation equation for $\xi^\mu$. On
the plane $\{n_1^\mu,n_2^\mu\}$ orthogonal to the line-of-sight ($n^a_\mu n_b^\mu =
\delta^a_{\;b}, n^a_\mu\bar{k}^\mu=0$), and setting $\xi_a(\lambda) = n^a_\mu
\xi^\mu(\lambda) \equiv \mathcal{D}^b_{\;a}(\lambda) \dot{\xi}_b(0)$, it turns out
\begin{equation}\label{eq:geodev}
\ddot{\mathcal{D}}^a_{\;b}=\mathcal{R}^a_{\;c}\mathcal{D}^c_{\;b}
\end{equation}
with initial conditions $\mathcal{D}^a_{\;b}(0) = 0, \dot{\mathcal{D}}^a_{\;b}(0) =
\delta^a_{\;b}$, the dot referring to a derivation with respect to $\lambda$. Here
$\mathcal{R}^a_{\;b}\equiv R^\alpha_{\;\mu\nu\beta}\bar{k}^\mu \bar{k}^\nu n_a^\alpha
n_b^\beta = - \frac{1}{2}R_{\mu\nu}\bar{k}^\mu \bar{k}^\nu \delta^a_{\;b} +
C^\alpha_{\;\;\mu\nu\beta}\bar{k}^\mu \bar{k}^\nu n_\alpha^a n_b^\beta$ is the optical tidal
matrix written in terms of the Riemann tensor $R^\alpha_{\;\mu\nu\beta}$ or in terms of the
Ricci and Weyl tensors, respectively $R_{\mu\nu}$ and $C^\alpha_{\;\mu\nu\beta}$. This
latter expression highlights the sources of the isotropic and anisotropic deformation of the
original image.

%In a perturbed universe with metric $g_{\mu\nu} = a^2(\eta)\bar{g}_{\mu\nu} = a^2(\eta)
%(\bar{g}^{\mathrm{RW}}_{\mu\nu} + h_{\mu\nu})$, \Eref{eq:geodev} is solved order-by-order in
In a perturbed universe with metric $g_{\mu\nu} = a^2(\eta) (\bar{g}^{\mathrm{RW}}_{\mu\nu}
+ h_{\mu\nu})$, \Eref{eq:geodev} is solved order-by-order in $h_{\mu\nu}$. %One can restrict
%to the metric $\bar{g}_{\mu\nu}$, since the Maxwell equations are not affected by a
%conformal transformation.
Here we consider scalar, vector, and tensor perturbations of the
Robertson-Walker metric allowing for curvature,
\begin{equation}\label{eq:metric}
\fl\qquad g_{\mu\nu}\rmd x^\mu\rmd x^\nu = a^2(\eta)\{
 -(1-2\phi)\rmd\eta^2 + 2B_i\rmd\eta\rmd x^i + [(1+2\psi)\gamma_{ij}+2E_{ij}]\rmd x^i\rmd x^j \},
\end{equation}
with $\nabla^i B_i = \nabla^i E_{ij} = E^i_i = 0$. The spatial metric $\gamma_{ij}\rmd
x^i\rmd x^j=\rmd\chi^2+S_K^2(\chi)\rmd\Omega^2$ is written in terms of the angular diameter
distance $S_K(\chi)=\sin(\sqrt{K}\chi)/\sqrt{K}$, $K=\{-1,0,1\}$, of the comoving radial
distance $\chi$, and of the infinitesimal solid angle $\rmd\Omega^2$. Exploiting the
conformal invariance of null geodesics, for $\bar{g}_{\mu\nu} = a^{-2}(\eta)g_{\mu\nu}$ the
optical tidal matrix reads
%At zeroth and first order, the optical tidal matrix reads
\begin{equation}
\fl\qquad\mathcal{R}^{a(0)}_{\;b}=-K\delta^a_{\;b}\;,\quad\mathcal{R}^{a(1)}_{\;b}=D^a
D_b(\phi+\psi+B_\c+E_{\c\c}) + K(E^a_{\;b}-E_{\c\c}\delta^a_{\;b}),
\end{equation}
$D_a$ being the covariant derivative with respect to the spatial metric $\gamma_{ij}$ and
$\c$ denoting the component along the line-of-sight. Accordingly, \Eref{eq:geodev} leads to
\begin{equation}\label{eq:opticaltidalmatrix}
%\fl\qquad
\ddot\mathcal{D}^{(0)}=-K\mathcal{D}^{(0)}\;,\quad
\ddot\mathcal{D}^{(1)}=-K\mathcal{D}^{(1)}+ \mathcal{R}^{(1)}\mathcal{D}^{(0)}.
\end{equation}

The solution $\mathcal{D}=\mathcal{D}^{(0)}+\mathcal{D}^{(1)}$ is finally rescaled by the
(angular) distance of the source galaxy, $d_\mathrm{A} (\lambda)$, to get the amplification
matrix $\mathcal{A}_{ab}=\mathcal{D}_{ab}(\lambda) /d_\mathrm{A}(\lambda)$; its diagonal and
off-diagonal terms, which account for the isotropic and anisotropic deformation of the
original image, are the observed quantities. Eventually, one has to integrate over the
distribution of sources $n(\chi)$ along the line-of-sight (notice that $\rmd\lambda =
a^2\rmd\chi$). Usually a fitting function of the form $n(z)\propto
(z/z_\s)^\alpha\exp[-(z/z_\s)^\beta]$ is taken to reproduce the observed distribution of
sources as a function of redshift $z$. Neglecting vector and tensor perturbations, from
$\mathcal{R}^{a(1)}_{\;b}\equiv D^a D_b\Phi$ the convergence field $\kappa =
(1-\Tr\mathcal{A})$ reads
\begin{equation}\label{kappa}
\fl\kappa(\btheta) = \int_0^{z_H}\rmd z\,n(z) \kappa(\btheta,z)
                = \int_0^{\chi_H} \rmd\chi\,n(\chi) \int_0^\chi \rmd \chi'\,
                   \frac{S_K(\chi)S_K(\chi'-\chi)}{S_K(\chi')}
                   \Delta_2\Phi[S_K(\chi')\btheta,\chi']\\
\end{equation}
where the deflecting potential $\Phi$ is calculated solving the field (e.g. Einstein)
equations.

In the flat sky approximation, the convergence and shear power spectra are
\begin{equation}\label{eq:Pk}
P_\kappa(\ell)=P_\gamma(\ell)=\frac{1}{4} \int \rmd\chi\,g^2(\chi)
                                 \left[ k^4 P_\Phi(k,\chi)\right]_{k=\ell/S_K(\chi)}
\end{equation}
where $g(\chi)=\int_\chi^{\chi_\mathrm{H}}\rmd\chi'\,n(\chi')S_K(\chi'-\chi)/S_K(\chi')$ and
$P_\Phi$ is the three-dimensional power spectrum of the deflecting potential. Two-point
correlation functions in the real space, like top-hat shear or aperture mass variances, are
filtered integrals of this quantity~\cite{wlreview}.

% ==========================================================================================

\section{Quintessence by cosmic shear: Parameterizations vs ``physical'' models}

%\subsection{Parameterizations vs ``physics''-inspired models}
The use of parameterizations for the quintessence equation of state generally assumes a
Friedmann-Robertson-Walker universe, thus excluding a priori other options for the dark
energy sector. Moreover, every parametrization is affected by the choice of a
dataset-dependent pivot redshift~\cite{HuJain04}, by the consistency with a model for the
speed of sound determining the formation of structure, and by the large number of parameters
required to suitably account for a ``realistic'' dynamics over a wide range of
redshift~\cite{corasaniti}. Dealing with ``physics''-inspired models one would overcome
these problems, aiming to investigate if a class of theory is compatible with observations
at low and high redshift.

We explore two ordinary quintessence scenarios, realized by a self-interacting scalar degree
of freedom $Q$ with Ratra-Peebles (RP) and supergravity (SUGRA) potentials
\begin{equation}\label{eq:RP-SUGRA}
V_\mathrm{RP}(Q)=M^{4+\alpha}/Q^\alpha;\quad
V_\mathrm{SUGRA}(Q)=M^{4+\alpha}\exp(Q^2/2M_\mathrm{Pl}^2)/Q^\alpha
\end{equation}
which guarantee the (partial) solution of the coincidence problem~\cite{EQ0}. The mass scale
$M$ is uniquely determined once $\alpha$ and the density parameter $\Omega_\q$ are fixed.

\begin{figure}[thf]
 \centering
   \includegraphics[width=7cm]{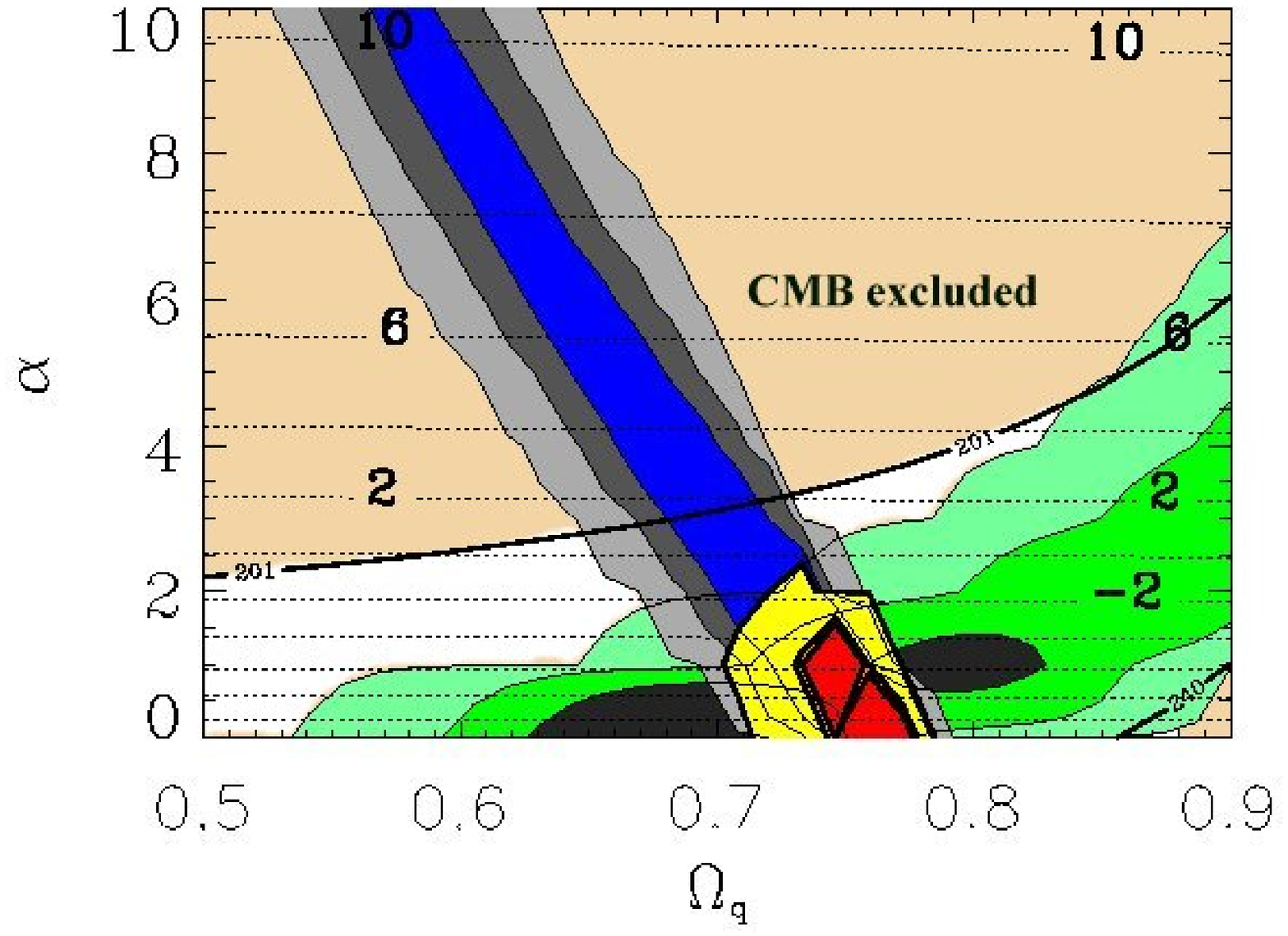}
   \includegraphics[width=7cm]{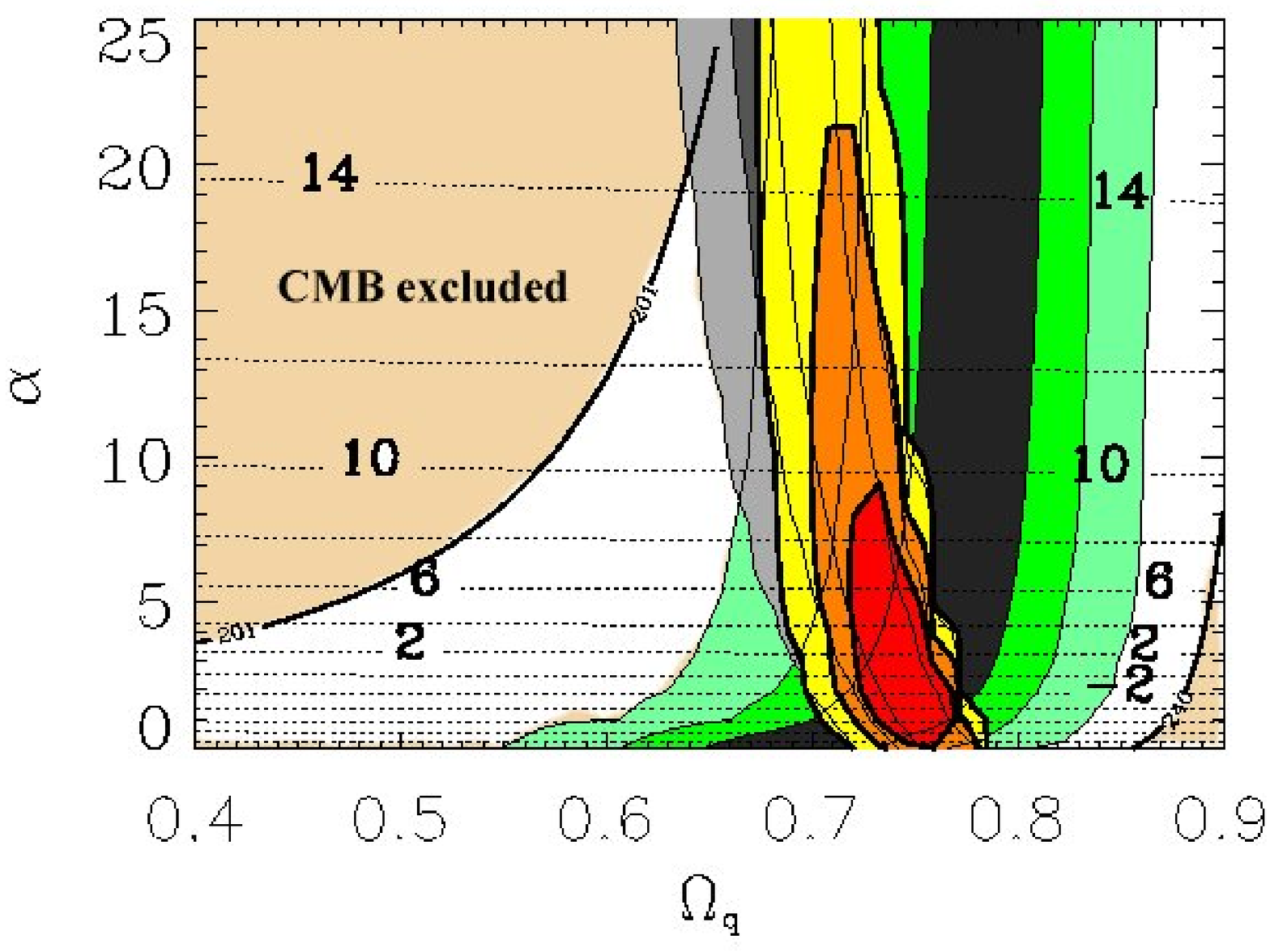}
  \caption{Likelihood analysis of the quintessence parameter space for RP (left) and SUGRA
  (right) models, marginalizing over $(n_s,z_s)$, using CFHT cosmic-shear data (blue contours,
  at 68, 95, and 99\%), the ``goldset'' of SnIa (green), and both jointly (red). The shaded
  region is excluded since it corresponds to a location of the first peak of CMB $C_\ell^\mathrm{TT}$
  not compatible with WMAP-3yr data (considering the very conservative range $201 \lesssim
  \ell_{1\mathrm{st}}\lesssim 240$). Dotted lines refer to $\log_{10}M/$GeV; see~\Eref{eq:RP-SUGRA}.}
 \label{fig:jointanalysis}
\end{figure}

Background and perturbations evolution in linear regime are computed solving the Einstein
and Klein-Gordon equations by means of a Boltzmann code described in~\cite{ru02}. Dealing
with general relativity, on sub-horizon scales one can safely use the Poisson equation to
relate the deflecting potential $P_\Phi=4P_\phi$ to the power spectrum of matter
perturbations. In order to account for the non-linear matter power spectrum, we use two
linear-to-non-linear mappings~\cite{LNLmappings} (see~\cite{sur04} a generalization in
scalar-tensor theories). Although calibrated on $\Lambda$CDM $N$-body simulations, they
provide a quite safe recipe also for QCDM models, provided the linear regime is properly
taken into account using the correct linear growth factor and the spectra normalized to
high-redshift (by CMB). Indeed, the $Q$-field mostly acts on the background dynamics
affecting the onset of the non-linear regime, while the successive evolution on small scales
(affecting the structure of single galaxies/halos, bias, etc.) is essentially dictated by
astrophysical processes.

% ==========================================================================================

\section{Joint cosmic-shear--SnIa data analysis}\label{sec:dataanalysis}

%\subsection{Cosmic shear alone}
We combined VIRMOS-Descart, CFHTLS-deep, and CFHTLS-wide/22~deg$^2$ top-hat shear variance
data (see~\cite{stum06} for references) to investigate the sensitivity to the description of
the non-linear regime~\cite{LNLmappings}. The results look different when using the
Peacock~\&~Dodds (1996) or the Smith \etal (2003) prescriptions, which are based on
different modeling of the non-linear clustering. Notice that both mappings do not include
the effects of baryons, not negligible on the scales of interest to cosmic
shear~\cite{baryons}. Interestingly, the quintessence parameter space $(\Omega_\q,\alpha)$
seems not to be sensitive to non-linear gravitational clustering, probing that quintessence
primarily acts on geometry.

%\subsection{Cosmic shear + SNe + CMB}
Combining the cosmic-shear data with the ``goldset'' of supernovae Ia (SnIa)~\cite{goldset},
\fref{fig:jointanalysis}, one can safely constraint RP models ($\alpha<1, \Omega_\q =
0.75^{+0.03}_{-0.04}$ at 95\%) because of the strong degeneracy between the two observables.
More care is needed for SUGRA models, where the superposition of likelihood contours is
severely dependent on reliability of their location, ultimately dependent on systematics.
Eventually, one can extract the value of the mass scale $M$, whose contours follow the
estimate $\log_{10}(M/\mbox{GeV}) \simeq(19\alpha-47+\log\Omega_\q)/(\alpha+4)$ which holds
as long as $Q\sim M_\mathrm{Pl}$. Actually, notice that redshift and shear calibration of
datasets are a crucial point~\cite{errcalibration}.

\begin{figure}[thf]
%\centering
   \includegraphics[width=7.0cm]{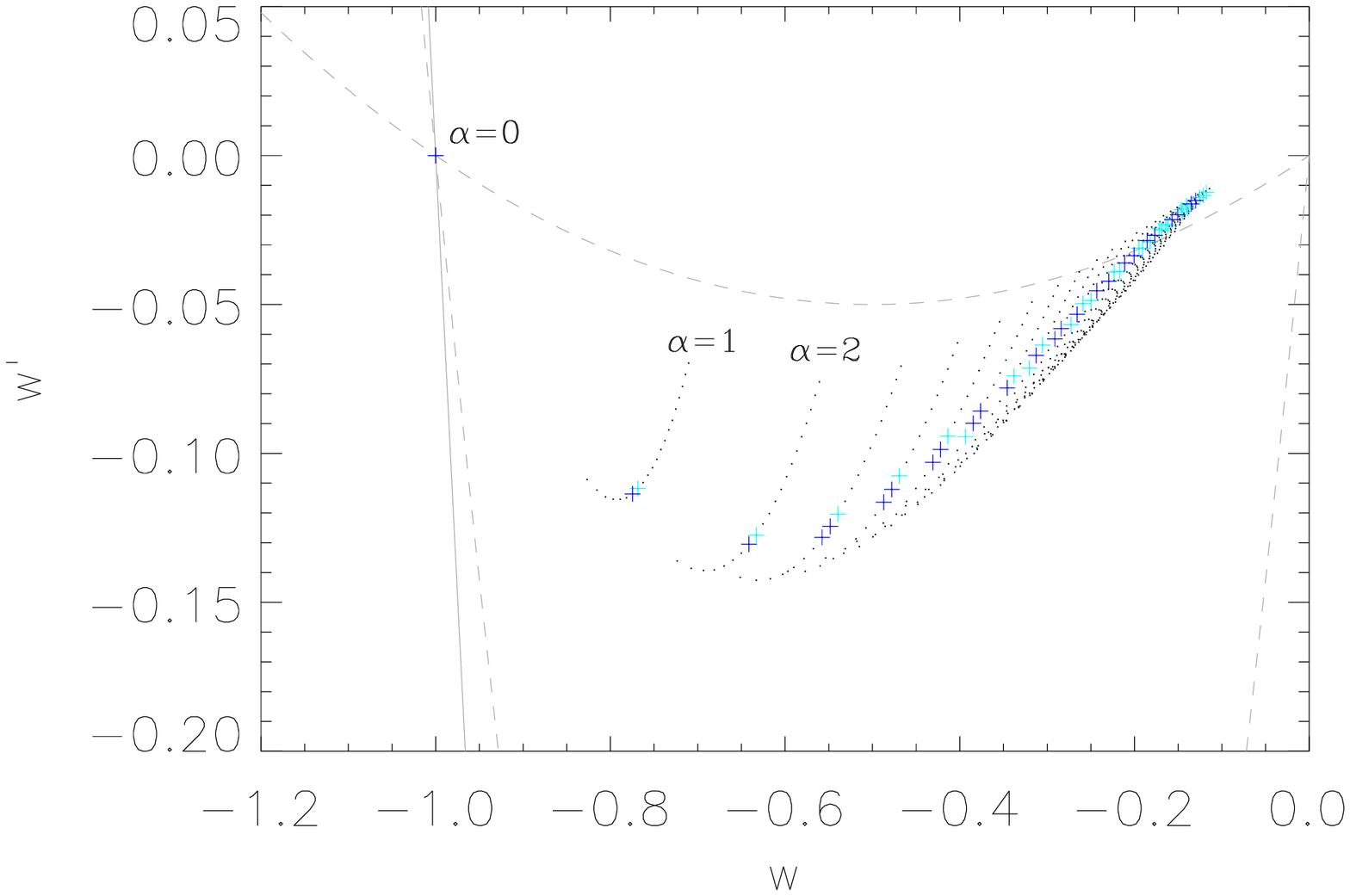}
   \includegraphics[width=7.0cm]{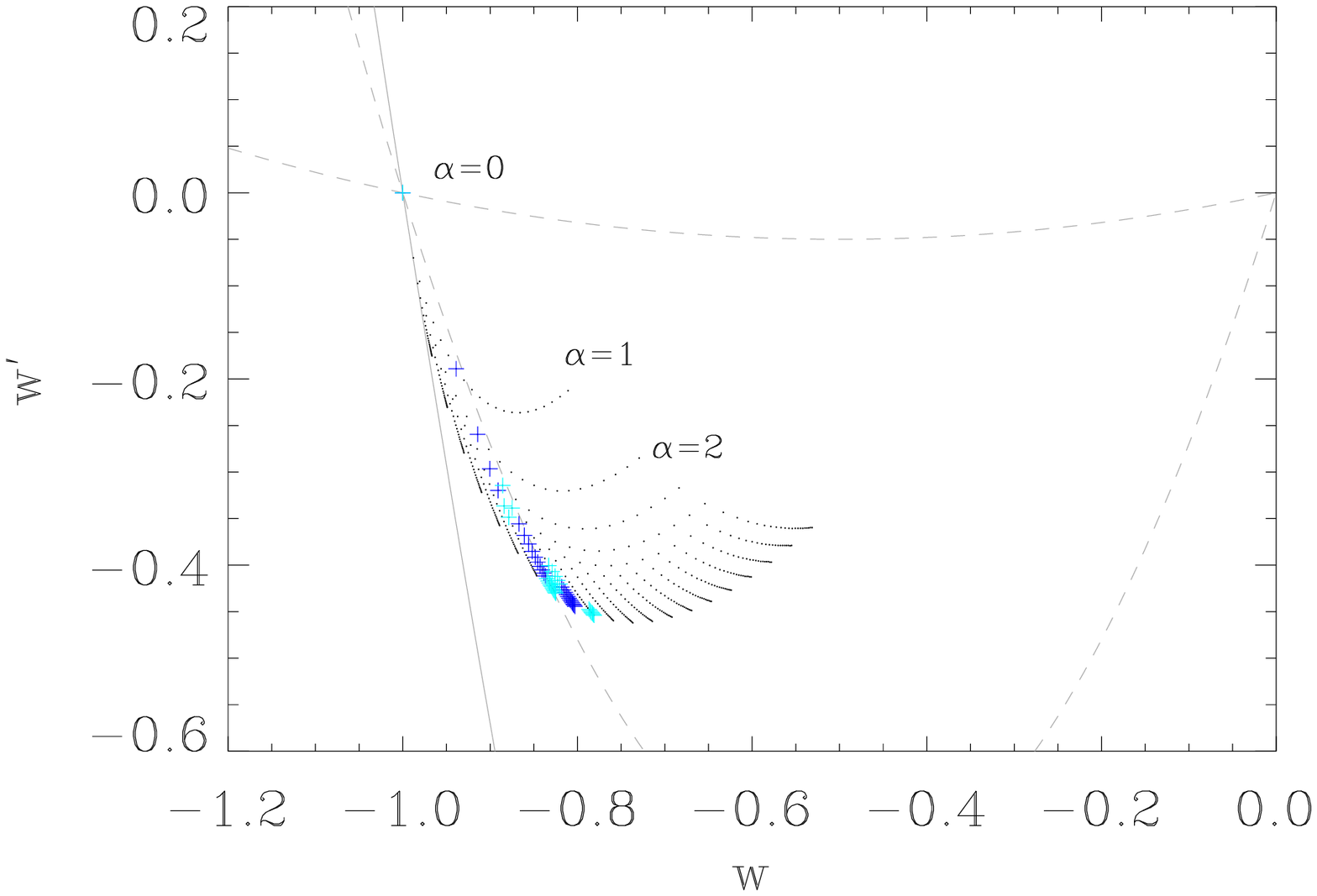}
  \caption{Likelihood analysis of the quintessence parameter space $(\Omega_\q,\alpha)$
   transposed in the $(w,w')$ parameter space (at $z=0$), for RP (left) and SUGRA (right) models,
   using only the combined cosmic shear data sets. Dark and light crosses refer to points
   lying respectively within the 68 and 95\% confidence levels of the original likelihood,
   while dots correspond to remaining points of the $(\Omega_\q,\alpha)$ grid. Solid and
   dashed lines correspond to the $c^2_\s<1$ constraint for the $Q$-field and the class
   of ``freezing'' quintessence model, respectively. The cosmological constant is recovered
   for $\alpha=0$.}
 \label{fig:w0wa}
\end{figure}

Exploiting the one-to-one relation between $(\Omega_\q,\alpha)$ and the values of the
quintessence equation of state and its time derivatives (valid for $\alpha\neq0$) at
whatever redshift, we translate the likelihood contours of the joint analysis in a $(w,w')$
parameter space ($w'\equiv\rmd w/\rmd\ln a$), evaluated at $z=0$; see \fref{fig:w0wa}. This
is useful to compare the models at hand with other classes of models~\cite{BargerScherrer},
and with the dynamical constraints $w'>-3(1-w^2)$ (solid line), corresponding to a
$Q$-field's speed of sound $c^2_\s<1$, and $3w(1+w)<w'<0.2w(1+w)$ (dashed line),
characterizing the so-called ``freezing'' models. For every set of points characterized by
the same $\alpha$ (linearly varying by $\Delta\alpha=1$), points span the $\Omega_\q$ range
$[0.4,0.9]$ from top to bottom (linear steps $\Delta\Omega_\q=0.025$).

% ==========================================================================================

\section{Wide surveys: forecasts}

\begin{figure}[thf]
 \centering
   \includegraphics[width=12cm]{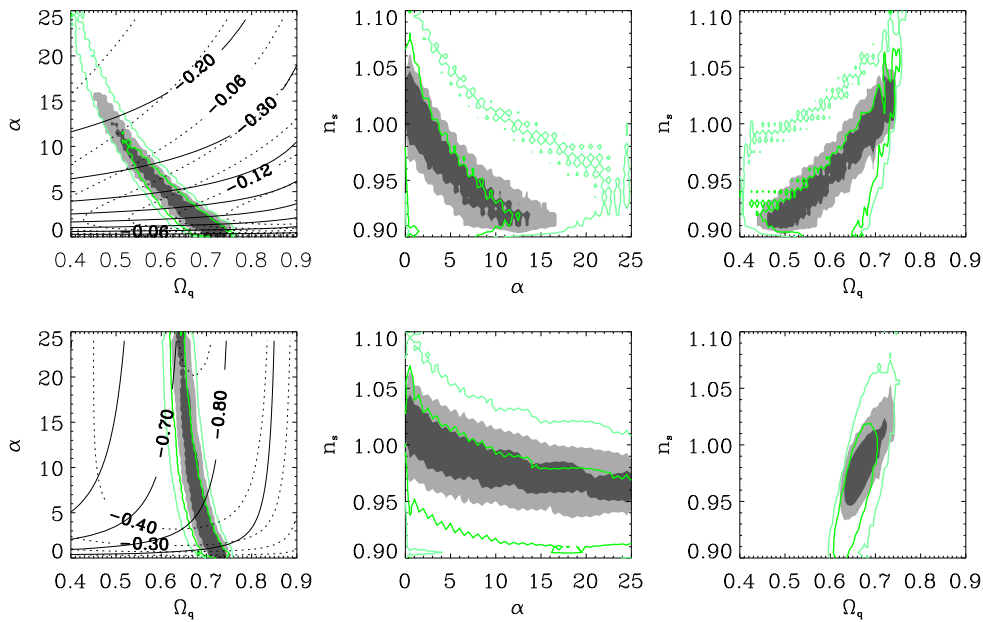}
  \caption{Likelihood analysis of the top-hat shear variance based on a synthetic realization
  of the full 170~deg$^2$ CFHTLS-wide survey (filled contours; 68 and 95\% c.l.) and considering
  only scales $\gtrsim 20$~arcmin (empty contours), for RP (upper) and SUGRA (lower). Solid
  (dotted) lines in the $(\Omega_q,\alpha)$ plane represent the value of $w$ ($w'$) at $z=0$.}
 \label{fig:wide}
\end{figure}

%\subsection{CFHTLS synthetic data analysis}
Focusing on wide surveys (\fref{fig:wide}), one can investigate angular scales where the
contamination of the non-linear regime is reduced (a residual being always present because
of the integration along the line-of-sight; see \Eref{kappa}). Using a synthetic realization
of the CFHTLS-wide full survey, we performed a likelihood analysis using only angular scales
$\gtrsim 20$~arcmin. Since it corresponds to cutting off larger wavemodes $k$, the contours
concerning all the cosmological parameters we considered broaden. However, the contours of
the quintessence parameter space are less affected by this cut. In particular, SUGRA models
seem to be very slightly dependent on $\alpha$ but highly sensitive on $\Omega_Q$. Finally,
remark that the results depend on $n(z)$, which we suppose to be the same of that measured
for the 22~deg$^2$ dataset.

\begin{figure}[thf]
 \centering
   \includegraphics[width=14cm]{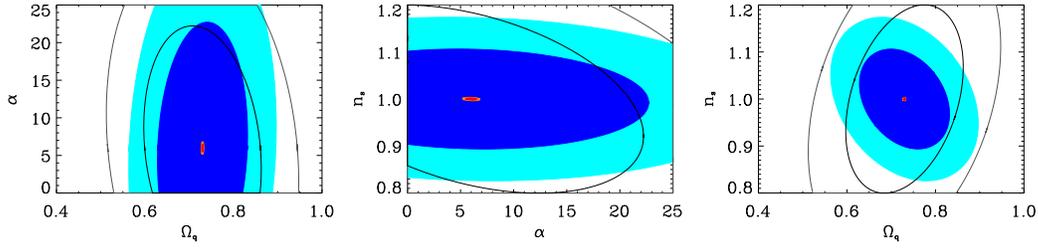}
  \caption{Fisher analysis forecast for a CFHTLS-wide 170~deg$^2$ survey (wide/blue ellipses) and
  a DUNE-like $20\,000$~deg$^2$ survey (small/red ellipses; contours at 1 and $2\sigma$) for the
  $(\Omega_\q,\alpha,n_\s)$ parameter space, marginalizing over $(\tau_\mathrm{reion},z_\s)$.
  Filled (empty) contours refer to SUGRA (RP) models (for RP, only CFHTLS-wide contours are shown),
  both centered at the same fiducial model.}
 \label{fig:DUNE}
\end{figure}

%\subsection{Forecasts for \DUNE}
%\DUNE \footnote{ {\tt www.dune-mission.net}} \cite{DUNE}
This analysis points towards the gain achievable by a very-wide survey, for which the
quality image requirements very likely need space-based observations. We thus forecast the
improvement on the RP and SUGRA models considering a DUNE-like mission~\cite{DUNE}, a
shallow survey covering $20\,000$~deg$^2$. The results depicted in \fref{fig:DUNE}
illustrate the gain with respect to the full CFHTLS-wide survey. It has to be stressed,
however, that this estimation is based on an approximate distribution of sources, assumes a
perfect correction of the point spread function, and finally is marginalized over a small
number of cosmological parameters.

% ==========================================================================================

\section{Concluding remarks}
Using a general formulation of weak-lensing valid for every metric theory of gravity, we
dispose of tools to study several classes of the dark energy sector at both low and high
redshift, avoiding the use of parameterizations which difficult a safe combination of data
sets and can hardly match any well-defined theory. We investigate two classes of ordinary
quintessence models by means of real cosmic shear data, and combining with supernovae data,
putting forward the interest of using CMB data to normalize the spectra and thus define the
onset of the non-linear regime of structures formation. The contamination of non-linear
gravitational clustering can be reduced when disposing of wide surveys, which reasonably
require space-based missions to achieve a high-precision characterization of the dark-energy
sector.

\ack The authors warmly thank Jean-Philippe Uzan, Yannick Mellier and Alexandre
R\'efr\'egier for fruitful discussions and comments on the manuscript.
 C.S. thanks IAP for hospitality.

% ==========================================================================================

\section*{References}
%\begin{thebibliography}{10}
%\bibitem{book1} Goosens M, Rahtz S and Mittelbach F 1997 {\it
%The \LaTeX\ Graphics Companion\/} (Reading, MA: Addison-Wesley)
%\bibitem{eps} Reckdahl K 1997 {\it Using Imported Graphics in \LaTeX\ } (search CTAN for the file `epslatex.pdf')
%\end{thebibliography}

\end{document}